\documentclass[iop]{emulateapj}

\usepackage{amsmath}

\newcommand{\Msun}{M_{\odot}}
\newcommand{\logZ}{[Fe/H]}

\shorttitle{SFH Random Uncertainties}
\shortauthors{Dolphin}

\begin{document}


\title{On the Estimation of Random Uncertainties of Star Formation Histories}


\author{Andrew E. Dolphin}
\affil{Raytheon Company, Tucson, AZ, 85734}
\email{adolphin@raytheon.com}

\begin{abstract}
The standard technique for measurement of random uncertainties of star formation histories (SFHs) is the bootstrap Monte Carlo, in which the color-magnitude diagram (CMD) is repeatedly resampled.  The variation in SFHs measured from the resampled CMDs is assumed to represent the random uncertainty in the SFH measured from the original data.  However, this technique systematically and significantly underestimates the uncertainties for times in which the measured star formation rate is low or zero, leading to overly (and incorrectly) high confidence in that measurement.  This study proposes an alternative technique, the Markov Chain Monte Carlo (MCMC), which samples the probability distribution of the parameters used in the original solution to directly estimate confidence intervals.  While the most commonly used MCMC algorithms are incapable of adequately sampling a probability distribution that can involve thousands of highly correlated dimensions, the Hybrid Monte Carlo algorithm is shown to be extremely effective and efficient for this particular task.  Several implementation details, such as the handling of implicit priors created by parameterization of the SFH, are discussed in detail.
\end{abstract}





\section{Introduction} \label{sec-intro}

Star formation histories (SFHs) of nearby galaxies can be estimated by analysis of their resolved stellar content, as many features present in a color-magnitude diagram (CMD) provide evidence of star formation at specific ages.  For example, blue or red supergiants provide evidence of recent star formation, while blue horizontal branch stars indicate the presence of ancient populations.  Relative strengths of age-related CMD features can be used to estimate the star formation rates (SFRs) at those ages, leading to an overall estimate of the SFH \citep{hod89}.

More quantitative methods for the measurement of SFHs have been introduced \citep[e.g.,][]{gal96,tol96,dol97,her99,hol99,har01}, and despite differences in implementation share a common overall approach.  First, synthetic CMDs are generated for a variety of potential SFHs through application of stellar evolution models, an observational error model (typically utilizing artificial star tests), and other parameters (e.g., distance, extinction, IMF, and unresolved binaries).  Second, the level of agreement between observed and synthetic CMDs is quantified using some goodness-of-fit metric, which is used to measure the SFH.

While significant work has been dedicated to the derivation and implementation of SFH measurement techniques, estimation of uncertainties in SFH measurements has been given considerably less attention.  On the topic of systematic uncertainties due to isochrone physics, \citet{dol12} proposed a technique for estimating the size of these uncertainties.  However, an understanding of random uncertainties is equally critical, especially when analyzing less-populated CMDs where random uncertainties are likely to dominate.  Examples of this include small regions within galaxies used to estimate ages of specific young populations such as those near a supernova \citep{mur11} or UV-bright regions \citep{sim13}.  Likewise, analysis of stars in individual clusters can be studied \citep[e.g.,][]{joh12,ham13}.  Finally, when making a comparative analysis of two systems \citep[e.g.,][]{wei13}, the relative uncertainties are dominated by random errors.  The present study evaluates the common approach used for this problem and several alternatives.

\section{Bootstrap Monte Carlo} \label{sec-bootstrap}

The bootstrap Monte Carlo technique is the standard method used in the estimation of random uncertainties in SFH measurements.  The approach involves three steps.  First, one generates an adequate number of resampled CMDs.  Assuming that the CMD has been binned into a Hess diagram, these can be easily generated by using values in the Hess diagram of either the original photometry or the best-fitting model as the means of Poisson distributions in each Hess diagram bin, and generating random deviates to create resampled photometry.  The SFH is then measured for each of the resampled CMDs.  Finally, the variation of recovered SFHs is calculated.  Provided a sufficiently large number of resamplings, that variation is assumed to equal the uncertainty of the original solution due to random errors.

\begin{figure*}
\epsscale{1.0}
\plottwo{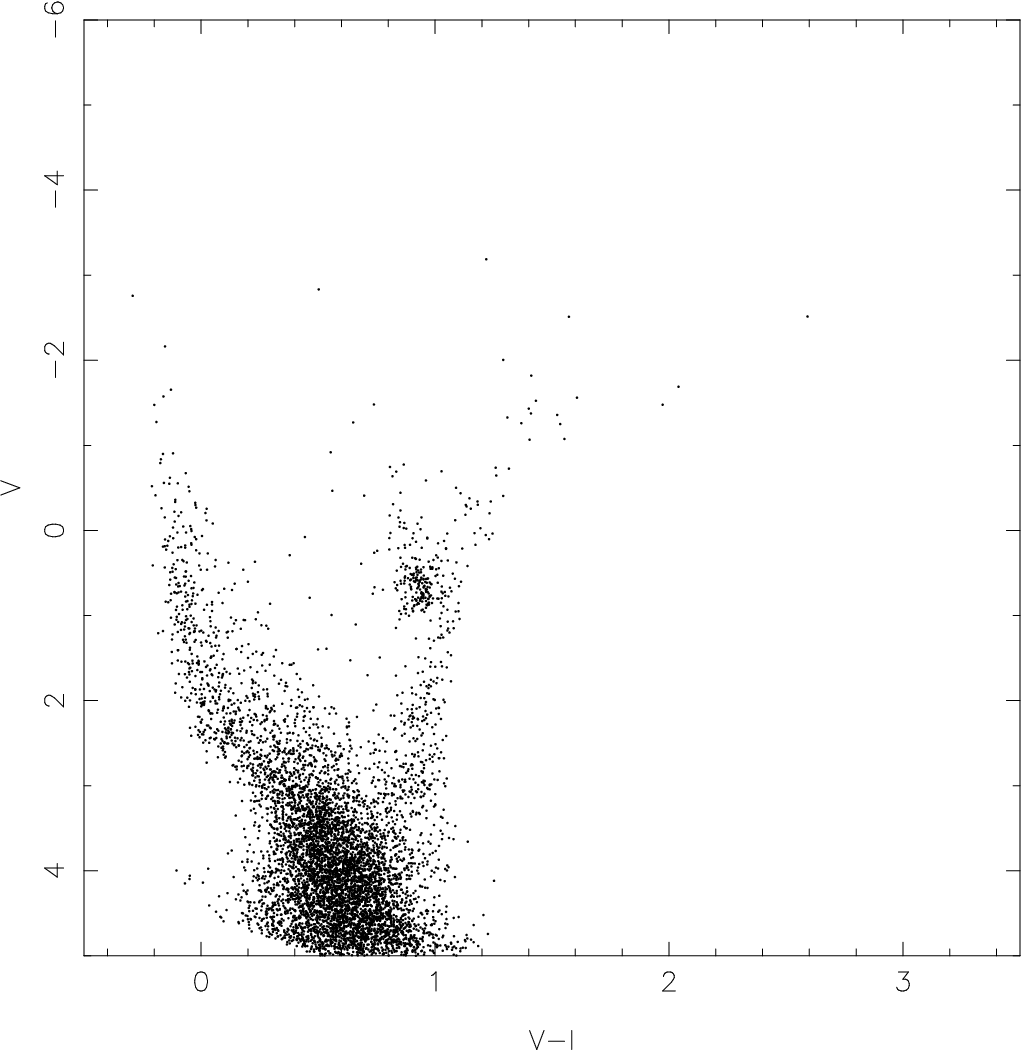}{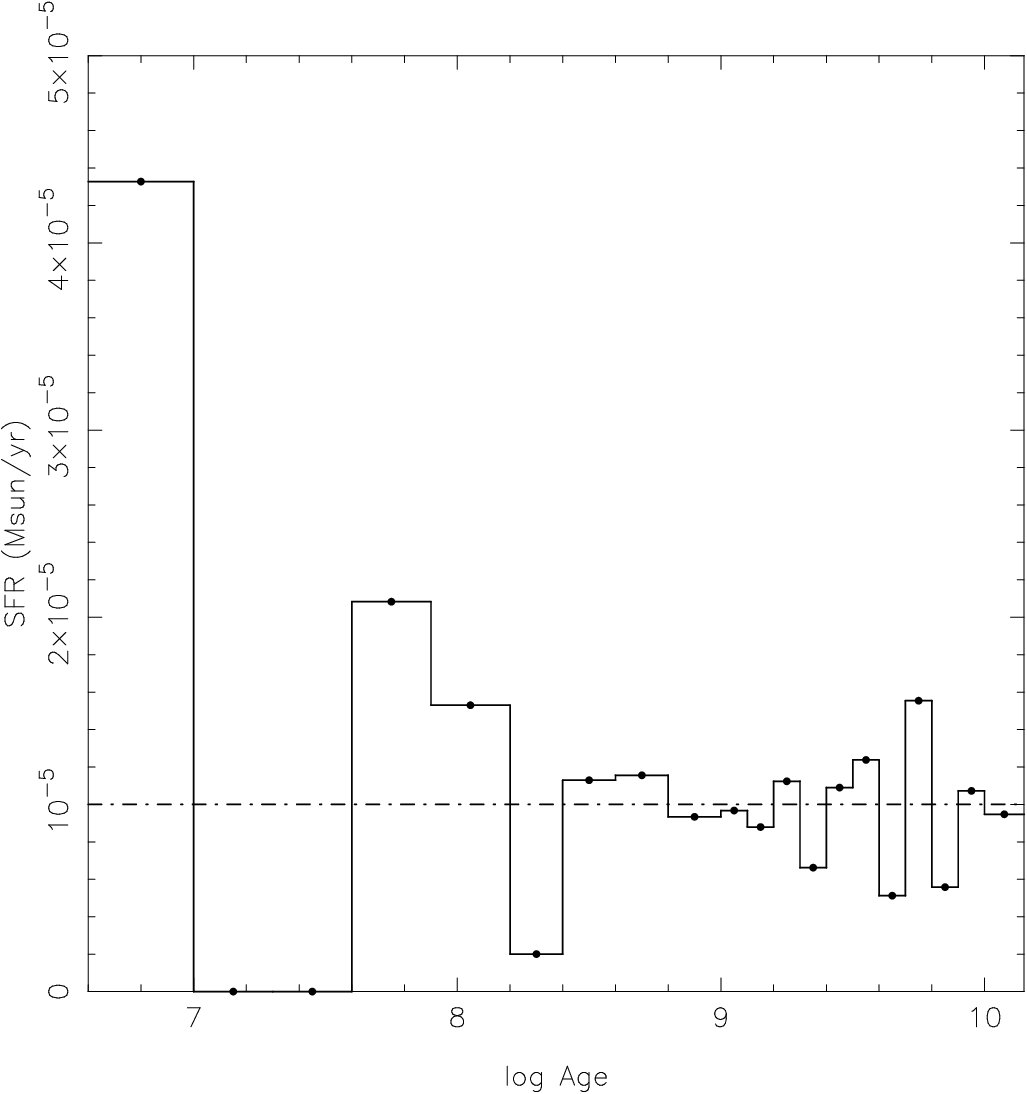}
\caption{Simulated color-magnitude diagram and maximum likelihood SFH.  The population was generated with a constant SFR of $10^{-5} \Msun yr^{-1}$ and metallicity of $\logZ=-0.45$.  In the right panel, the measured SFH is denoted with solid lines, while the true SFH is denoted with a dash-dot line.  Differences between true and measured SFH are due entirely to random errors created by stochastic sampling of the CMD. \label{fig-sample1}}
\end{figure*}

\begin{figure*}
\epsscale{1.0}
\plottwo{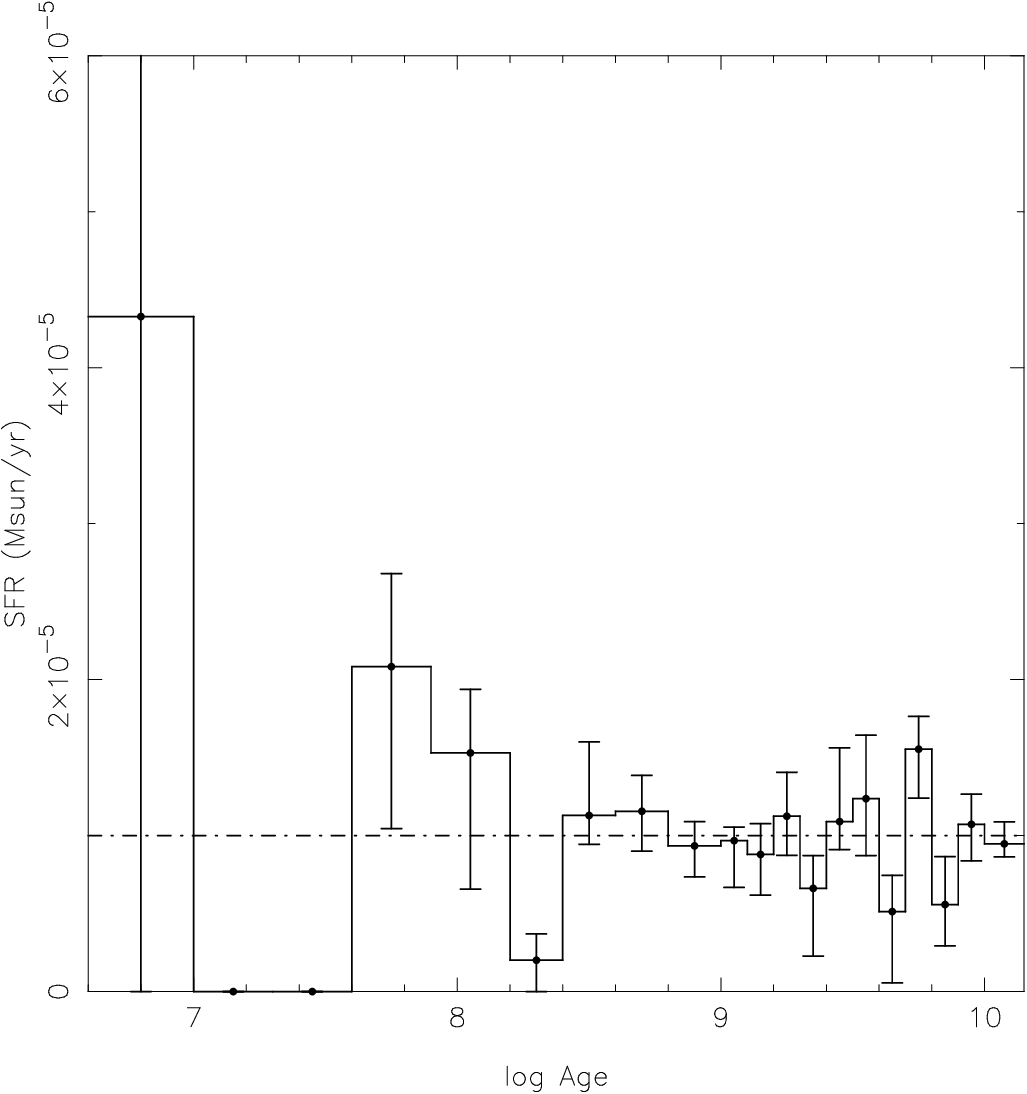}{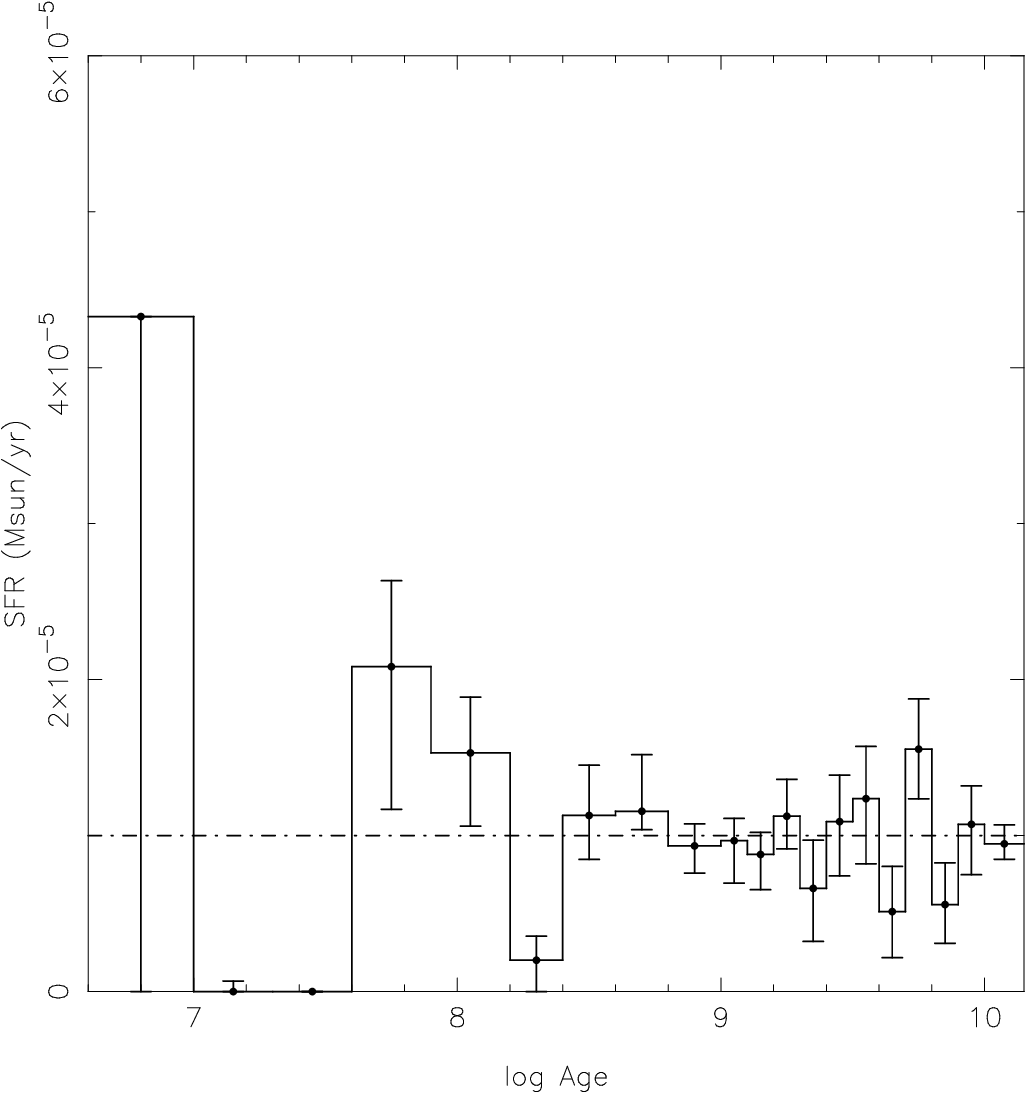}
\caption{Measured SFH for the CMD shown in Figure \ref{fig-sample1}, with uncertainties estimated using bootstrap Monte Carlo techniques.  Uncertainties in the left panel are computed by resampling the original photometry; those in the right panel are computed by resampling the best-fitting model.  In both panels, the true SFR is denoted with a dash-dot line.  Note that, in both cases, the measured uncertainty is zero for the two bins between log age of 7.0 to 7.6 (10 to 40 Myr). \label{fig-bootstrap}}
\end{figure*}

To illustrate this approach, Figure \ref{fig-sample1} shows a synthetic population generated with constant SFR and constant metallicity, and the best-fit SFH as determined by MATCH \citep{dol02}.  Two sets of bootstrap Monte Carlo runs were made, one by resampling the original photometry and the other by resampling the best fitting synthetic CMD.  Uncertainties estimated with the two variants are shown in Figure \ref{fig-bootstrap}.

In both versions of the bootstrap Monte Carlo, time bins for which zero star formation was detected in the original solution showed zero star formation in most or all of the Monte Carlo runs.  The reason for this is due to stochastic sampling of age-specific indicators.  If no core helium-burning stars corresponding to 10-40 Myr old populations are present (as was the case in this example), those populations are unlikely to be measured in the resulting SFH.  Consequently, data created by resampling either the observed or model CMD will also fail to have those stars indicating a 10-40 Myr population, causing the Monte Carlo runs to return zero star formation in that age range.

This example illustrates a significant limitation of bootstrap Monte Carlo techniques.  Given that the CMD in Figure \ref{fig-sample1} was generated using a constant SFR and that no other errors were introduced (e.g., isochrone sets, photometric error model, etc.), any differences between the input and measured SFHs is due solely to random errors created by stochastic sampling of the CMD.  Therefore, any estimate of those random errors must show that the measured SFH is consistent with a constant SFR.  Since the bootstrap Monte Carlo results violate this requirement, the assumption that best-fit SFHs of resampled CMDs represent the confidence limits on the original solution to the observed data is invalid.  Instead, one needs to measure confidence intervals in that original solution.

\section{Measuring the Probability Density Function} \label{sec-pdf}

The direct approach for determining a confidence interval for a quantity of interest is to calculate the probability density as a function of that quantity, and determine a probability density threshold above which the desired percent of the probability lies.  The confidence interval is the region with probability density above that threshold.  For example, the $68\%$ confidence interval of a normal distribution defined by a mean $\mu$ and standard deviation $\sigma$ is bounded by $\mu-\sigma$ and $\mu+\sigma$.

Applying the above approach to the measurement of SFHs, the probability density must be written a function of the SFH.  Assuming that the CMD is converted into a Hess diagram by binning, the probability of a model with a predicted $\lambda_i$ stars producing an observed $k_i$ stars in the $i^{th}$ Hess diagram bin is given by
\begin{equation}\label{eq-logP}
P(k_i|\lambda_i) = \frac{e^{-\lambda_i} \lambda_i^{k_i}}{k_i!}.
\end{equation}
The predicted number of stars in that bin is determined by the star formation history.
\begin{equation}\label{eq-lambda}
\lambda_i = \sum_j SFR_j v_{i,j},
\end{equation}
where $SFR_j$ is the SFR in time bin $j$, and $v_{i,j}$ is the number of stars that would be expected to fall within the $i^{th}$ Hess diagram bin, given that $SFR_j=1$ and all other times have zero star formation.

Since equations \ref{eq-logP} and \ref{eq-lambda} calculate the probability of randomly creating an observed Hess diagram given a SFH, Bayes' theorem allows the determination of the probability of the SFH given the observed Hess diagram:
\begin{equation}\label{eq-Bayes}
\begin{split}
P(SFR_1,...,SFR_N|k_1,...,k_M) \propto \\
P(k_1,...,k_M|SFR_1,...,SFR_N) \prod_j P(SFR_j),
\end{split}
\end{equation}
where $P(SFR_j)$ is the prior on the SFR in time bin $j$.  Unless there are external constraints on the SFH, the priors should be non-informative.  A uniform prior $P(SFR_j)=1$ is adopted here.

Once the probability density has been measured over an evenly spaced and well sampled grid of the SFHs of interest, the probability density for the SFR in time bin $j$ is calculated by marginalizing over all other variables, and the confidence limits can be computed.

The above approach works well for small numbers of dimensions, such as star clusters that can be treated as single stellar populations (SSPs).  In the SSP case, there is only a single SFR being measured, plus possibly a foreground/background star model scale factor.  It is thus feasible to measure the probability density as a function of several input parameters, such as age, metallicity, distance, extinction, IMF slope, and/or unresolved binary fraction \citep{ham13}.

For the case of galaxy SFHs, however, a typical solution has dozens of time bins.  (The examples presented in section \ref{sec-bootstrap} were solved with 20 time bins.)  Worse, if the entire population box is being measured directly, this is multiplied by dozens of metallicity bins and the space of all possible SFHs can easily exceed a thousand dimensions.  Even the case of a star cluster with prolonged star formation is likely to have of order ten age and/or metallicity bins, making direct sampling of the full PDF problematic.  So, while the measurement of the probability density over a several-dimensional space is possible and has been demonstrated effective for single stellar populations, a different approach is required if one wishes to estimate random errors of SFH measurements of mixed populations.

\section{Markov Chain Monte Carlo} \label{sec-markov}

\subsection{Overview} \label{subsec-mcmc-overview}

An alternative to numerical integration of the PDF is Markov Chain Monte Carlo (MCMC), a process designed to generate a set of samples whose density is proportional to the probability density.  The advantage of this approach is its efficiency: if implemented well, one can avoid computationally expensive measurement of the PDF in regions of zero probability.  This is especially helpful in a correlated space such as that seen in SFH solutions, where similarities between models corresponding to adjacent age or metallicity bins can create near degeneracies in their respective SFRs.

Several techniques for generating the MCMC samples exist, but share two important similarities.  First is that the samples are computed by use of a Markov Chain, a random process in which calculation of the next sample requires knowledge only of the current sample (i.e., no information of past samples is used).  The second is that transitions must obey the principle of detailed balance.  Denoting $P(s,s')$ as the probability for transitioning from state $s$ to $s'$, and $f(s)$ as the probability density at state $s$, detailed balance requires that
\begin{equation}\label{eq-balance}
f(s) P(s,s') = f(s') P(s',s).
\end{equation}

Once the MCMC samples have been created, it can be used to identify confidence intervals.  A simple solution is to adopt the mean and standard deviation, but this will only identify the region of highest probability if the probability density function is symmetric about its maximum.  A better approach is to identify the confidence limit bounding the region of highest probability, making use of the fact that the density of MCMC samples is proportional to the probability density.  It follows that the confidence interval is the narrowest interval containing the desired fraction of the samples.  For example, to compute the $68\%$ confidence interval, one could determine the interval bounded by the $0^{th}$ and $68^{th}$ percentiles, the $1^{st}$ and $69^{th}$, and so on and select the narrowest of those.

\subsection{MCMC Sampling Techniques} \label{subsec-mcmc-techniques}

The most common algorithm used to create MCMC samples is the Metropolis-Hastings algorithm \citep{met53,has70} or various adaptations of that algorithm.  The algorithm uses a two-step process to determine the next sample.  The first step is the calculation of a proposed new state $s'$ based on the current state $s$.  The proposal function must be symmetric, in that the probability of proposing a transition from $s$ to $s'$ must equal the probability of proposing the reverse transition.  This is commonly accomplished by setting $s'$ equal to $s$ plus a random number drawn from a zero-mean multivariate normal distribution.

The second step is determining the probability of acceptance of the new state.  If the probability density at the new state equals or exceeds that of the old state, the probability of acceptance is 1.  If not, it equals the ratio of the probability densities, $f(s') / f(s)$.

The challenge in implementing an efficient Metropolis-Hastings algorithm in a space with large dimensions is to create a function that can generate proposals reasonably far from the current state while also having a reasonably high acceptance rate.  For the case of SFH measurement, a solution to this was not found.  Either the step size was too small so that the entire probability space was not sampled, or the acceptance rate was too low to generate an acceptable number of independent samples.  Other common algorithms, such as slice sampling \citep{nea03} or Gibbs sampling \citep{gem84}, also proved unsuitable for this application.

The hybrid Monte Carlo (HMC) algorithm \citep{dua87} solves the problem of creating proposals with large step size and high acceptance rate by using Hamiltonian evolution between states.  If the potential is set equal to $- \ln f(s)$; the initial velocity is randomly generated using a zero-mean, unity sigma normal distribution (in each dimension); and the numerical propagation uses an exactly reversible process, the acceptance probability equals the ratio of the final to initial Hamiltonian.  Since the Hamiltonian is conserved unless large step sizes cause truncation error, acceptance probability of near one can be achieved.

One challenge of adapting the HMC algorithm to the SFH problem is that the acceleration is equal to the gradient of the logarithm of the probability density, so any discontinuity is likely to cause truncation error in the propagation and reduce the acceptance rate.  The requirement that SFRs be non-negative sets up a potential discontinuity; the recommended solution is to reparameterize the SFR as the square of the parameter being propagated.  Due to sampling a space that is not linear in SFH, this implicitly introduces a prior of $1/\sqrt{SFH}$ on the sampled distribution that must be eliminated by weighting the samples by $\sqrt{SFH}$ when processing the samples.  An alternative approach that preserves the uniform prior is to set the SFR equal to the absolute value of the parameter being propagated.  This removes the possibility of negative SFRs, but leaves a discontinuity in the gradient's derivative and significantly reduces the acceptance rate.

\begin{figure}
\epsscale{1.0}
\plotone{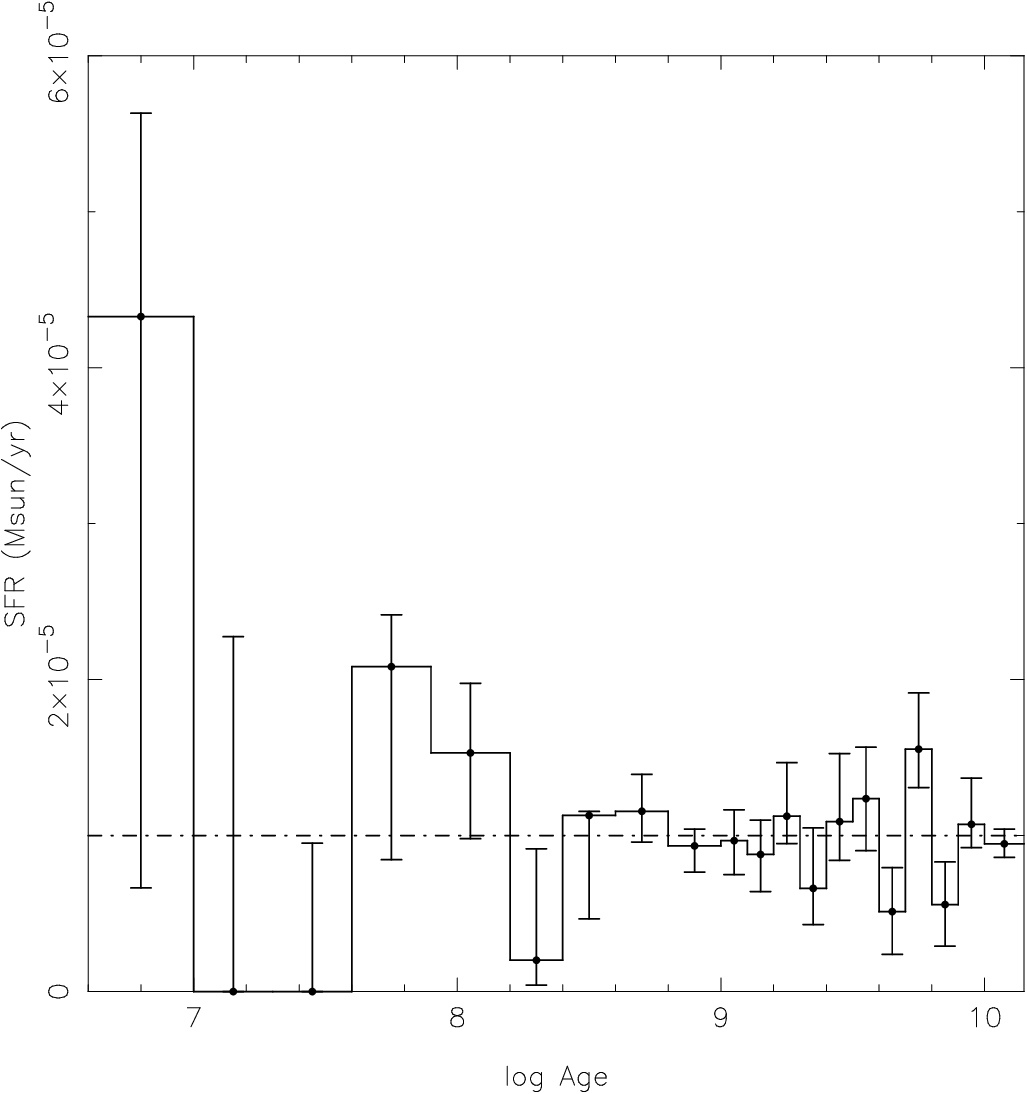}
\caption{Measured SFH for the CMD shown in Figure \ref{fig-sample1}, with uncertainties estimated using the HMC technique.  The true SFR is denoted with a dash-dot line.  Note that, unlike the bootstrap Monte Carlo (Figure \ref{fig-bootstrap}), non-zero uncertainties are estimated for the two bins between log age of 7.0 to 7.6. In addition, the upper error bar for the bin centered at 8.3 is much closer to truth than was the case in Figure \ref{fig-bootstrap}. \label{fig-hmc}}
\end{figure}

The uncertainties computed from the HMC routine are shown in Figure \ref{fig-hmc}.  For age bins in which the measured SFR was non-zero in the original solution, the uncertainties are very similar between the HMC and bootstrap MC.  However, the HMC technique is able to successfully measure uncertainties in the two bins with zero measured SFR.  A figure of merit is that the input (truth) SFR should be outside of the plotted 68\% confidence limits for about 32\% (6) of the twenty SFR measurements in Figure \ref{fig-hmc}.  This example had four points exceeding the limit, which is slightly lower but not a statistically significant discrepancy.

\subsection{Combining Independent Parameters} \label{subsec-mcmc-nozinc}

The preceding discussion has assumed that the parameters being sampled in the HMC process are the parameters one wishes to constrain (SFR vs. time).  However, SFH measurements from photometry deeper than the main sequence turnoff frequently measure the full population box of SFR vs. time and metallicity \citep[e.g.,][]{wei13}.  In this case, SFR as a function of time is calculated by summing across the metallicities.  The effect of summing $N$ values, each of which has a $1/\sqrt{x}$ prior, is to create a prior of $SFR^{N/2-1}$ on the total SFR.

A direct solution is to parameterize the SFRs so that the prior of the sum of $N$ independent values is uniform.  This can be achieved by defining the SFR as $\lvert x \rvert ^N$.  If $N$ is small, this approach can be effective. As it becomes larger (examples shown in this paper have 24 metallicity bins), the derivative of the probability with respect to $x$ can become very large, requiring exceedingly small propagation step sizes and thus long execution times.

An alternative solution is to use a nested HMC approach.  The main loop executes normally, using the same $SFR = x^2$ parameterization recommended in the previous section.  Each sample from this HMC is then used to determine the metallicity distribution as a function of age and as the initial solution for an HMC solution in which only the SFR vs. time is varied.  Given adequate burn-in (which, given the efficiency of the HMC approach, can be of order 10 samples), a sample can be obtained that has the desired prior on the total SFR at each time.  For $N=24$, this approach is an order of magnitude faster than that from the previous paragraph and produces equivalent error estimates.

\section{Other Parameters with Uncertainties} \label{sec-distance}

\begin{figure*}
\epsscale{1.0}
\plottwo{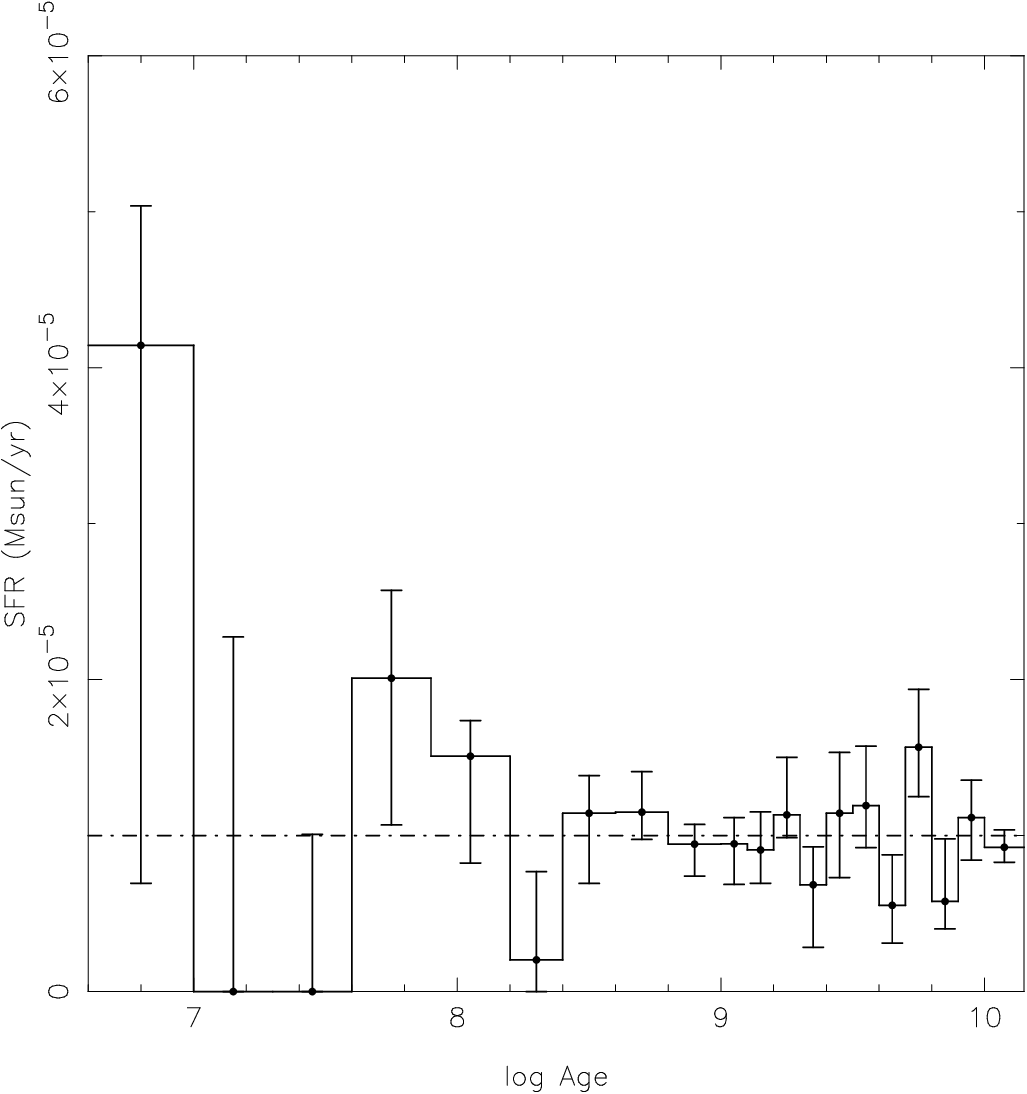}{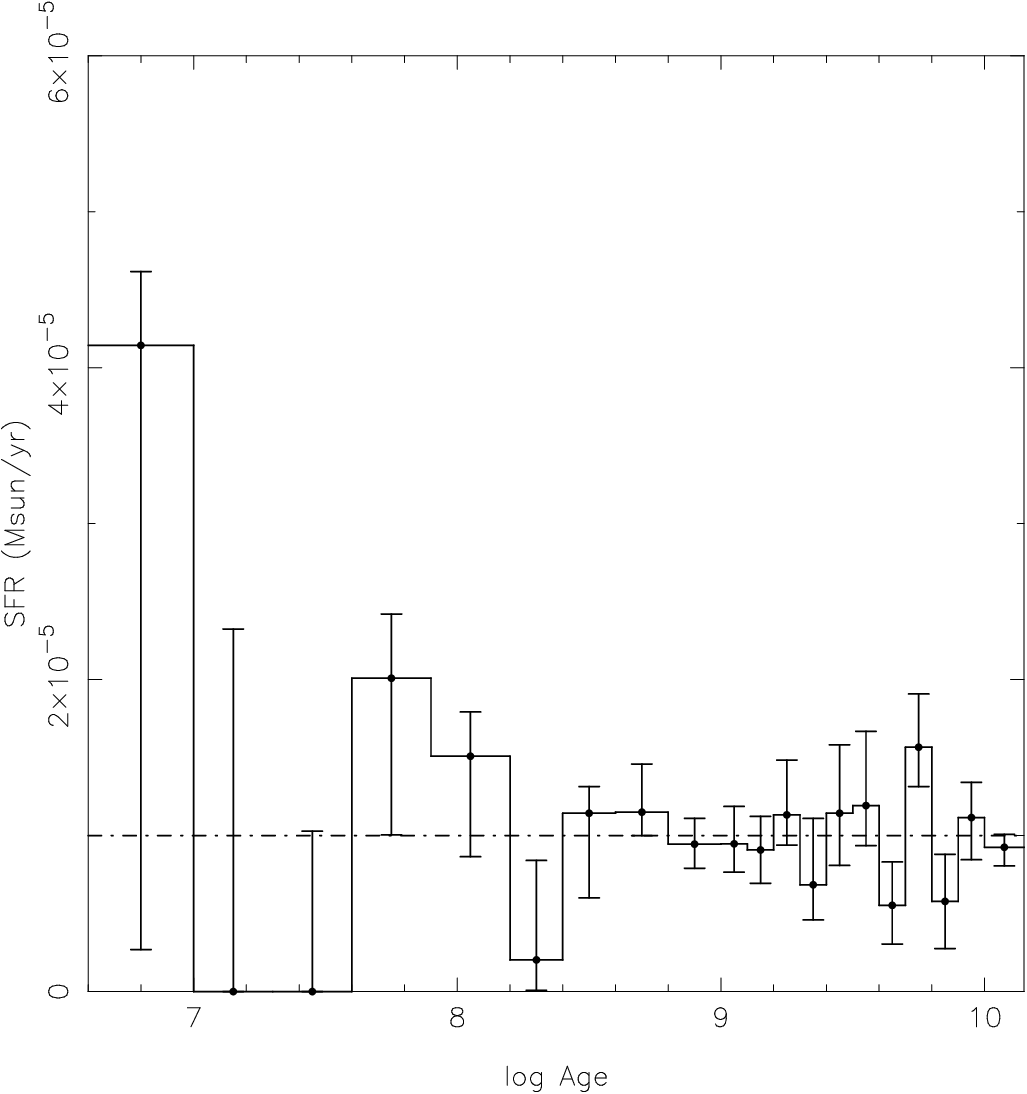}
\caption{Measured SFH, with uncertainties due to Poisson noise and uncertainties in distance and extinction.  In the left panel, the total uncertainty is calculated by running several HMC runs, with the number of samples of each proportional to the probability density in distance and extinction.  In the right panel, the total uncertainty is calculated by separately calculating Poisson uncertainty from a single HMC at the best-fitting distance and extinction, and distance/extinction uncertainty from variations in the best-fit SFH at the various distance and extinction points used.  Note that the overall estimated uncertainty is roughly the same, indicating that the two approaches produce equivalent results.  In both panels, the true SFR is denoted with a dash-dot line. \ref{fig-bootstrap}. \label{fig-distance}}
\end{figure*}

In addition to effects of Poisson noise, uncertainties on key assumed parameters (e.g., distance, extinction, IMF, and unresolved binaries) can introduce uncertainty in the SFH.  Ideally, these dependencies would be measured during the HMC process described in the previous section.  However, while the gradient of the probability as a function of SFR can be trivially calculated from a single set of basis functions, these parameters affect the basis functions themselves, making analytical evaluation of the gradients challenging.

Instead, a process similar to that outlined in Section \ref{sec-pdf} is recommended.  Assuming that the maximum likelihood is proportional to the probability marginalized over SFRs, each point in the distance/extinction/etc. space can be assigned a probability density proportional to the maximum likelihood of the SFH solution at that point.  Any external constraints on the parameters can be used as a prior.

The effects on the SFH uncertainties can be quantified precisely by executing an HMC test at each of these points, producing a number of samples proportional to the posterior probability density.  The final MCMC sample is then the combination of each of these.  A faster alternative is to compute statistics separately on the maximum likelihood solutions at each point, then add the uncertainties in quadrature to those from a single HMC run with the parameters corresponding to the maximum likelihood.  Figure \ref{fig-distance} shows total uncertainties (random plus those from distance and extinction uncertainty) calculated both ways; the two approaches produce equivalent uncertainties.

\section{Discussion} \label{sec-combined}

Combined with the technique for estimation of uncertainties due to isochrone uncertainties \citep{dol12}, the technique presented in this study allow one to robustly and accurately estimate the complete uncertainty of SFH measurements.  To help understand the relative impacts of the different sources of uncertainty, four simulated populations are analyzed covering the combinations of shallow and deep photometry, and few vs. many stars.  As before, all were generated with a constant SFR and metallicity.

\begin{figure*}
\epsscale{1.0}
\plotone{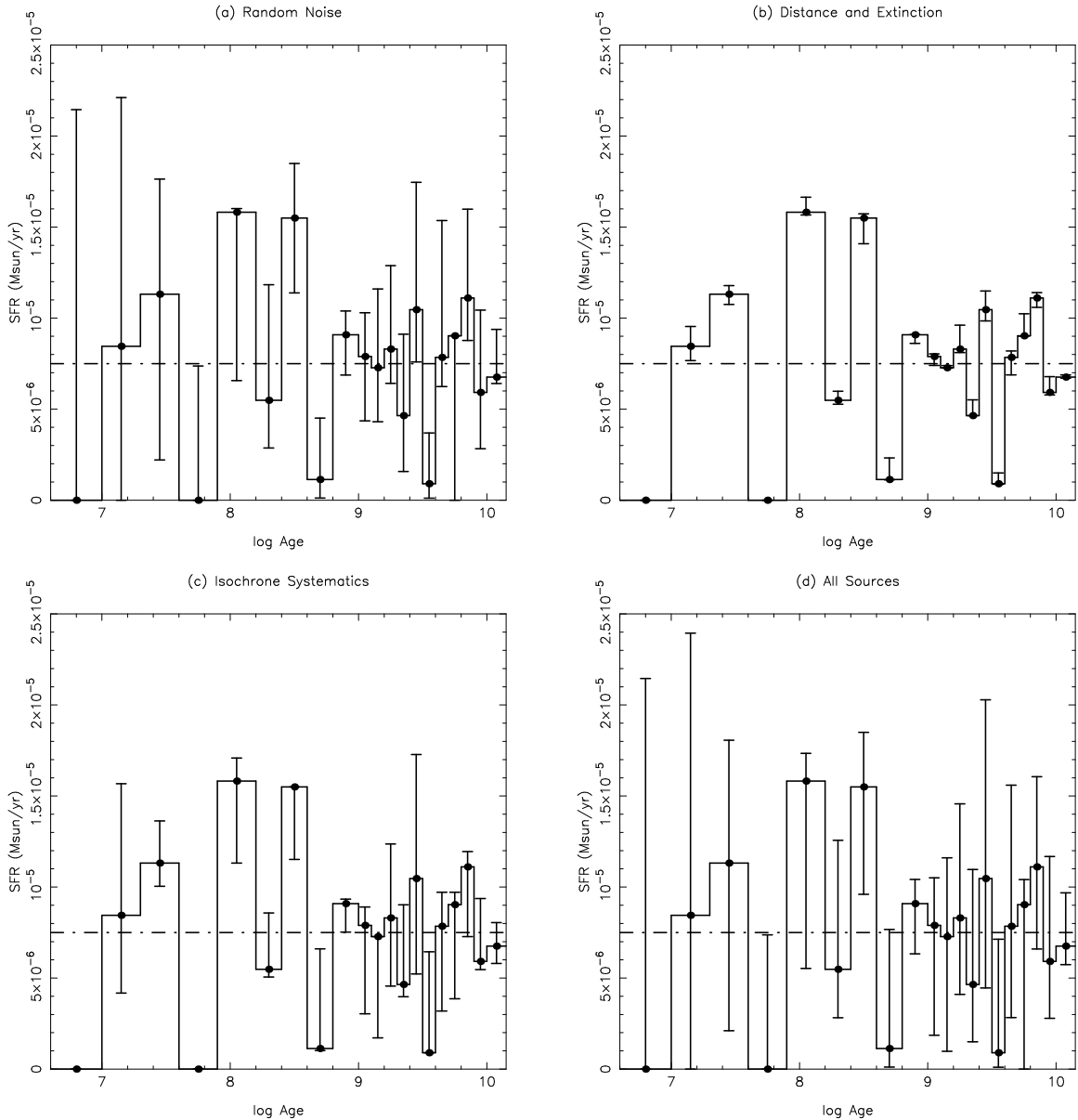}
\caption{Measured SFR of a simulated system with photometry complete to $M_V = 5$ and $\sim 5000$ observed stars.  The panels show uncertainties from three distinct sources: (a) Poisson noise, (b) distance and extinction uncertainty, and (c) systematic isochrone uncertainty.  The last panel (d) shows the combined error, calculated by adding the three individual components in quadrature. \label{fig-deep1}}
\end{figure*}

\begin{figure*}
\epsscale{1.0}
\plotone{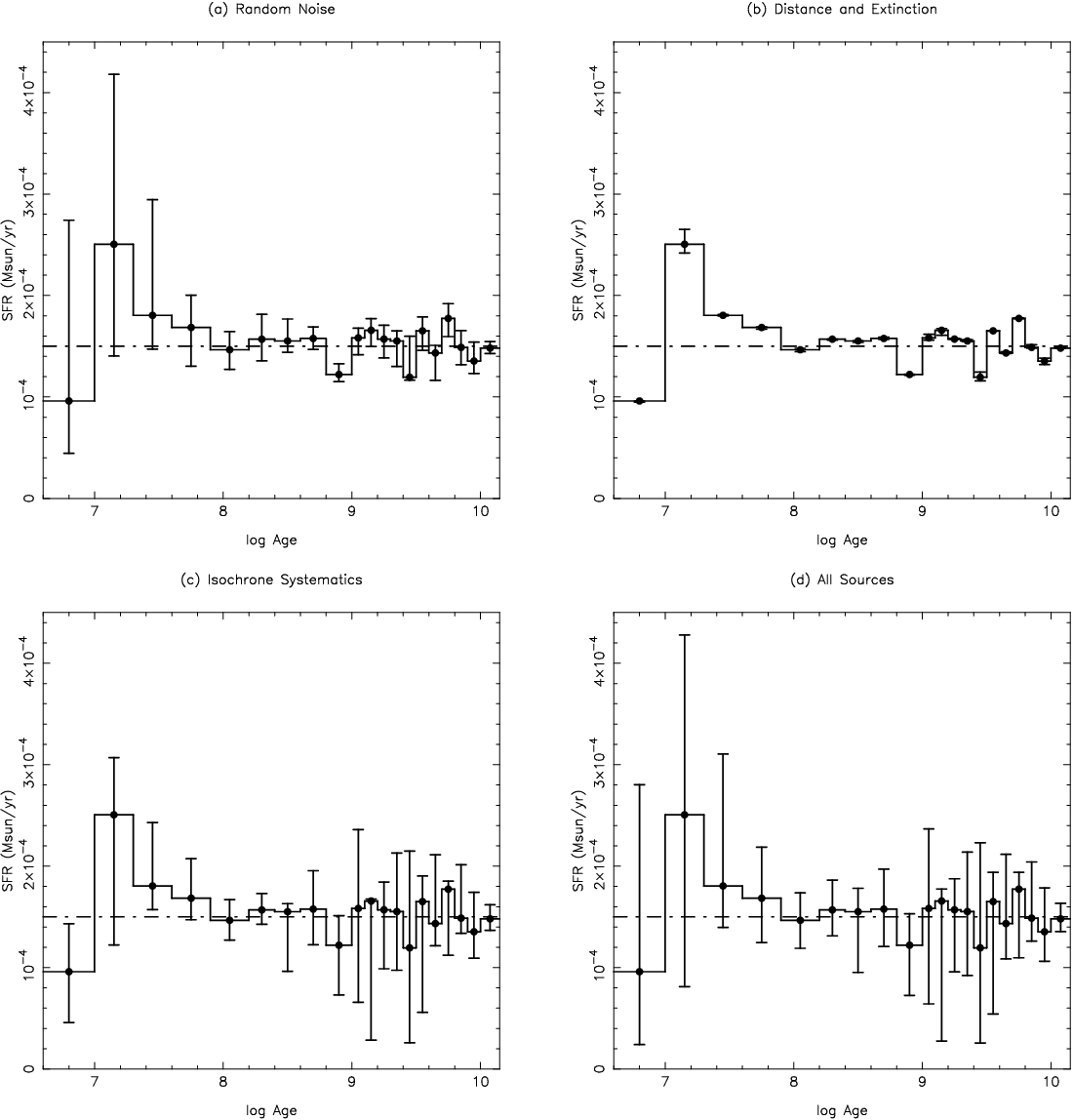}
\caption{Measured SFR of a simulated system with photometry complete to $M_V = 5$ and $\sim 100,000$ observed stars.  Panels are the same as in Figure \ref{fig-deep1} \label{fig-deep2}}
\end{figure*}

Figure \ref{fig-deep1} shows contributions to the error budget for each of the three sources of uncertainty for a system with deep photometry but only $\sim 5000$ observed stars.  For the populations with deep photometry, the SFH was solved as a function of both age and metallicity as a several hundred parameter problem and the HMC algorithm was executed as described in Section \ref{subsec-mcmc-nozinc}.  In this example, a comparison of panels (a) and (d) shows that the total uncertainty is dominated by random errors.  Figure \ref{fig-deep2} shows equivalent results for a system with the same photometry depth but 20 times more stars.  Here, random errors dominate at young ages ($< \sim 10^8$ yr), systematics dominate at older ages.

\begin{figure*}
\epsscale{1.0}
\plotone{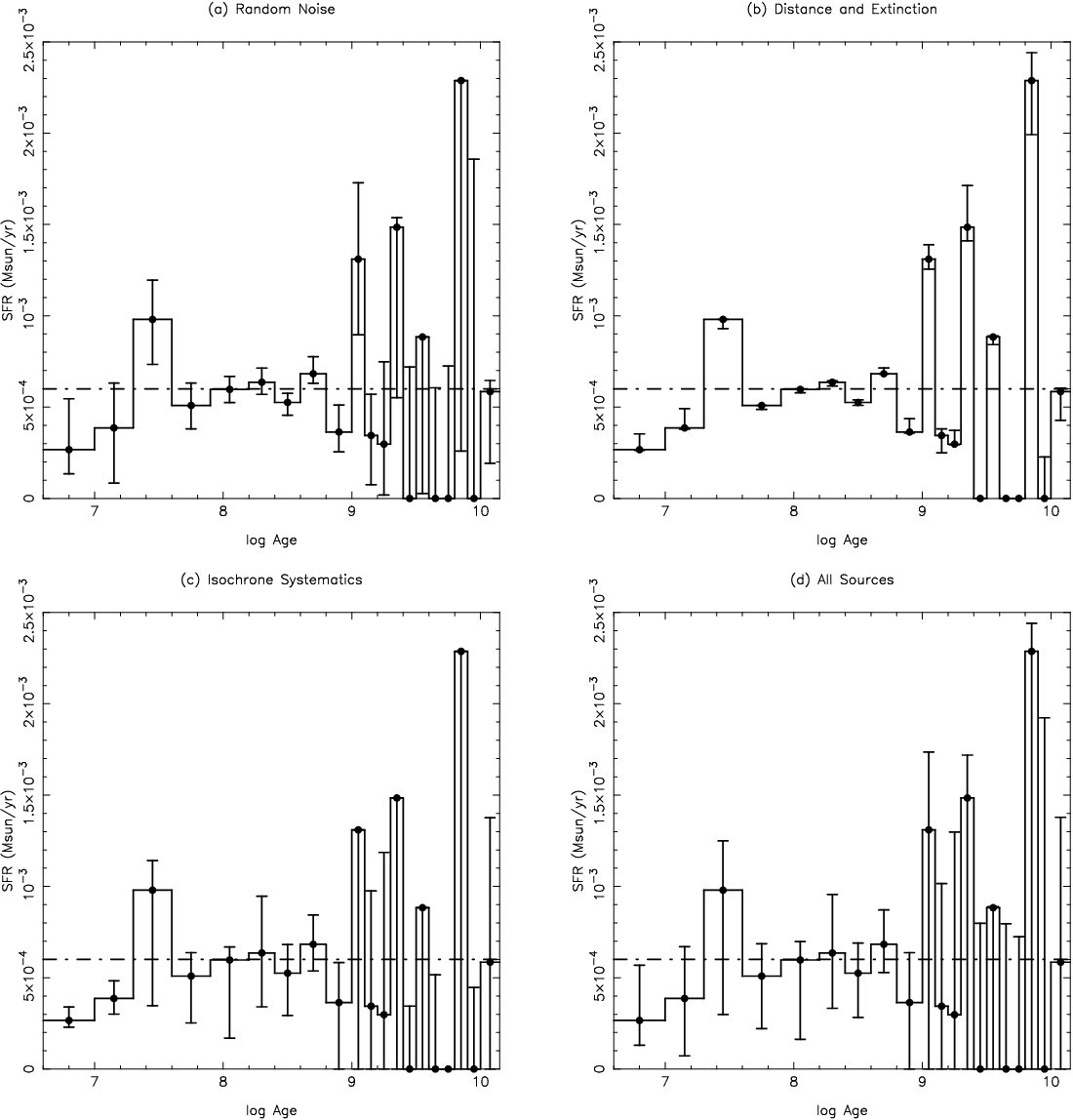}
\caption{Measured SFR of a simulated system with photometry complete to $M_V = 0$ and $\sim 5000$ observed stars.  Panels are the same as in Figure \ref{fig-deep1} \label{fig-shallow1}}
\end{figure*}

\begin{figure*}
\epsscale{1.0}
\plotone{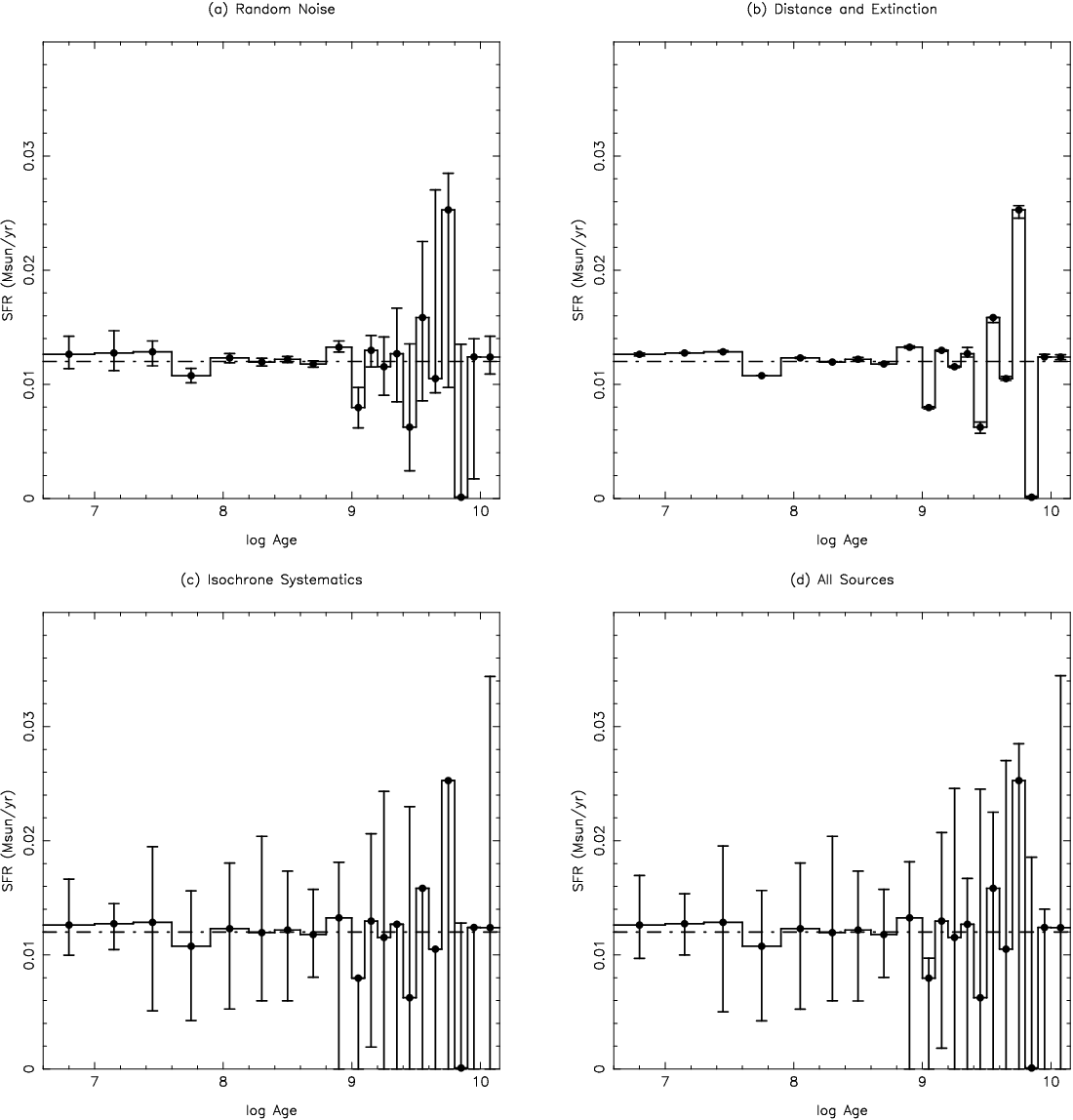}
\caption{Measured SFR of a simulated system with photometry complete to $M_V = 0$ and $\sim 100,000$ observed stars.  Panels are the same as in Figure \ref{fig-deep1} \label{fig-shallow2}}
\end{figure*}

Figures \ref{fig-shallow1} and \ref{fig-shallow2} show equivalent analysis for the case of photometry reaching only $M_V = 0$ (i.e., not reaching the horizontal branch or red clump).  For this, mean metallicity vs. time was assumed to follow one of three potential functional forms and was constrained to prevent metallicity from decreasing with time.  (As implemented in MATCH, the three forms are linearly decreasing $\logZ$ with age, linearly decreasing Z with age, and a form in which most enrichment happens at high age.)  Solutions and HMC runs with many potential chemical enrichment histories were made to determine the best solution and quantify the uncertainties.  In the case with few stars (Figure \ref{fig-shallow1}), both random and systematic error sources provide significant contributions, but uncertainties given a shallow, well-populated CMD (Figure \ref{fig-shallow2}) are dominated by systematic uncertainties throughout.

It is worth noting that in no case is uncertainty due to distance or extinction significant; this is due to the fact that the absolute magnitude and effective temperature shifts used to quantify systematic uncertainties greatly exceed the distance or extinction uncertainties.

\section{Summary} \label{sec-summary}

This paper examines the problem of estimating random uncertainties of SFHs.  Under inspection, the traditional process for this estimation (bootstrap Monte Carlo) is shown to significantly underestimate uncertainties in populations for which the measured SFH was zero.  The effect is also observed, but less dramatic, in populations for which the measured SFH was non-zero but significantly underestimated.  This is seen to be a result of particular indicator populations (e.g., blue helium burning stars for young populations) not being present in the photometry and thus not being created in the resampled photometry either.  The cause of this is that the set of best-fitting solutions to resampled photometry does not trace the probability density of solutions to the original photometry.

While marginalization of the probability density to determine confidence intervals would be a direct approach to solving this problem, the large number of dimensions and highly correlated space makes such an approach intractable for something beyond a single stellar population.  Instead, Markov Chain Monte Carlo (MCMC) techniques can be used to create a set of samples whose density is proportional to the probability density function.

Of the various MCMC algorithms, the hybrid Monte Carlo (HMC) algorithm works extremely efficiently in the high-dimensional space of SFH solutions, given a well-behaved gradient of the logarithm of the likelihood.  An effective parameterization that ensures the well-behaved gradient defines the SFR as the square of the parameter being solved; this eliminates both the boundary at zero SFR and discontinuity in the gradient at that location.  The only drawback is that weighting of the MCMC sample by $\sqrt{SFR}$ is required to eliminate the implicit prior created by that parameterization.

The case of measuring the population box directly (solving for the SFR across a two-dimensional grid of age and metallicity) is discussed.  Here, when marginalizing the samples over metallicity to estimate the SFR vs. time only, the prior is significantly affected.  An approach for resolving this problem is suggested, in which MCMC samples from the full solution are used to seed SFR-only HMC runs.

Incorporation of uncertainties in other fitting parameters, such as distance, extinction, IMF slope, or unresolved binary populations, is also discussed.  Because these are generally insignificant when compared with random and isochrone systematic uncertainties, a shortcut of treating the maximum likelihood solutions at each value as a weighted sample can be adopted without degrading the accuracy of the final solution.

Finally, random uncertainties estimated using the MCMC process are compared with isochrone systematic uncertainties estimated using the process developed by \citet{dol12}.  Random uncertainties are more significant for younger populations, deeper photometry, and CMDs with fewer stars.  This comparison also illustrates how uncertainties from all three error sources (random, isochrone systematic, and distance/extinction uncertainty) can be combined to determine the full uncertainty of the SFH.

\acknowledgments

Support for program number HST-GO-12055.15 was provided by NASA through a grant from the Space Telescope Science Institute, which is operated by the Association of Universities for Research in Astronomy, Incorporated, under NASA contract NAS5-26555.

\end{document}